\documentclass[12pt]{article}

\usepackage{amsmath,amssymb,graphicx} 
\usepackage{epsf}
\usepackage{pstricks}

\newcommand{\beq}{\begin{eqnarray}}
\newcommand{\eeq}{\end{eqnarray}}

\newcommand{\centeron}[2]{{\setbox0=\hbox{#1}\setbox1=\hbox{#2}\ifdim
\wd1>\wd0\kern.5\wd1\kern-.5\wd0\fi \copy0
\kern-.5\wd0\kern-.5\wd1\copy1\ifdim\wd0>\wd1
                                    \kern.5\wd0\kern-.5\wd1\fi}}
\newcommand{\ltap}{\>\centeron{\raise.35ex\hbox{$<$}}
                            {\lower.65ex\hbox{$\sim$}}\>}
\newcommand{\gtap}{\>\centeron{\raise.35ex\hbox{$>$}}
                            {\lower.65ex\hbox{$\sim$}}\>}

\newcommand\ZZ{\hbox{\zfont Z\kern-.4emZ}}
\font\zfont = cmss10 

\newcommand{\fref}[1]{fig.\ \ref{f.#1}}
\newcommand{\eref}[1]{eq.\ (\ref{e.#1})}
\newcommand{\erefn}[1]{ (\ref{e.#1})}

\newcommand{\cref}[1]{Chapter \ref{c.#1}}

\def\nn{\nonumber \\}

\def\beq{\begin{equation}}
\def\eeq{\end{equation}}
\newcommand{\ba}{\begin{array}}
\newcommand{\ea}{\end{array}}
\newcommand{\bea}{\begin{eqnarray}}
\newcommand{\eea}{\end{eqnarray} }
\newcommand{\bal}{\begin{align}}
\newcommand{\eal}{\end{align}}
\def\bi{\begin{itemize}}
\def\ei{\end{itemize}}
\def\ben{\begin{enumerate}}
\def\een{\end{enumerate}}
\def\beq{\begin{equation}}
\def\eeq{\end{equation}}
\def\bc{\begin{center}}
\def\ec{\end{center}}
\def\bt{\begin{table}}
\def\et{\end{table}}
\def\btb{\begin{tabular}}
\def\etb{\end{tabular}}
\newcommand{\bvec}{\left ( \ba{c}}
\newcommand{\evec}{\ea \right )}

\def\gev{\, {\rm GeV}}

\def\mass2{mass${}^2$}

\def\ra{\rangle}
\def\la{\langle}

\def\simlt{\stackrel{<}{{}_\sim}}
\def\simgt{\stackrel{>}{{}_\sim}}

\def\eps{\epsilon}

\begin{document}
\begin{titlepage}

\vskip1.5cm
\begin{center}
{\huge \bf Electroweak Precision Observables \vspace{.4cm} \\ and the Unhiggs}
\vspace*{0.1cm}
\end{center}
\vskip0.2cm

\begin{center}
{\bf Adam Falkowski$^{a}$ and Manuel P\'erez-Victoria $^{b}$}

\end{center}
\vskip 8pt

\begin{center}
$^a$ {\it NHETC and Department of Physics and Astronomy \\ 
Rutgers University, Piscataway, NJ 08855–0849, USA } \\
\vspace*{0.3cm}
$^b$ {\it CAFPE and Departamento de F\'{\i}sica Te\'orica y del Cosmos, \\ 
             Universidad de Granada, E-18071, Spain } 

\vspace*{0.3cm}

{\tt falkowski@physics.rutgers.edu, mpv@ugr.es}
\end{center}

\vglue 0.3truecm

\begin{abstract}
\vskip 3pt \noindent 

We compute one-loop corrections to the S and T parameters in the Unhiggs scenario.
In that scenario, the Standard Model Higgs is replaced by a non-local object, called the Unhiggs, whose spectral function displays a continuum above the mass gap.  
The Unhiggs propagator has effectively the same UV properties as the Standard Model Higgs propagator, which implies that loop corrections to the electroweak precision observables are finite and calculable.   
We show that the Unhiggs is consistent with electroweak precision tests when its mass gap is at the weak scale; in fact, it then mimics a light SM Higgs boson. 
We also argue that the Unhiggs, while being perfectly visible to electroweak precision observables, is invisible to detection at LEP.

\end{abstract}

\end{titlepage}


The Higgs boson is the last building block of the Standard Model (SM) that resists experimental verification.
Yet the whole theoretical and experimental consistency of the SM hinges on that element. 
The vacuum expectation value (vev) of the Higgs field breaks the electroweak symmetry, giving mass to the W and Z gauge bosons and the fermions. 
The tree-level exchange of the Higgs particle contributes to the scattering amplitude of longitudinally polarized electroweak gauge bosons in a way that makes the total amplitude consistent with unitarity, provided the Higgs mass is not larger than 1 TeV \cite{et}. 
The minimal Higgs sector with one Higgs doublet is automatically consistent with experimental data on flavor changing neutral currents and CP violation. 
Finally, radiative corrections from the Higgs boson affect the electroweak precision observables, notably the Peskin-Takeuchi S and T parameters \cite{PT}, and once again they are consistent with experiment, provided the Higgs boson mass is smaller than 145 GeV (at $2 \sigma$ confidence level \cite{gfitter}).  
Actually, the best fit is achieved for the Higgs boson mass somewhat smaller than the 115 GeV lower limit from LEP, which is probably the only source of tension within the minimal Higgs paradigm.    

In spite of this phenomenological success, theoretical arguments and prejudices (the hierarchy problem, in the first place) prompt searching for alternatives of the SM Higgs.
Surprisingly, finding a satisfactory alternative turns out to be highly non-trivial, 
and all proposed examples so far face more or less severe problems.      
Either new contributions to electroweak precision observables are unacceptably large (Higgsless), or one has to accept fine-tuning at least at the one-percent level (MSSM, Little Higgs, pseudo-Goldstone Higgs), or complicated ad-hoc theoretical structures have to be added. 
Moreover, extensions of the SM typically face flavor and CP problems, and/or a host of model-specific problems.  
In view of that, it is reasonable to ask if there exist unexplored theoretical directions that could present new model-building opportunities.    

Recently, ref.\ \cite{ST} proposed an interesting deformation of the SM Higgs. 
The general idea is to replace an elementary Higgs field with a more complicated object described by a non-local action.
In the model of ref.\ \cite{ST}, the kinetic term in momentum space is a non-analytic function of the form $(-p^2+\mu^2)^{2-d}$. 
The corresponding spectral density behaves as $\rho(s) \sim \sin (\pi d) s^{d - 2}$ in the UV.\footnote{One could also represent the Unhiggs as a large or infinite number of densely spaced Higgs bosons, each carrying a fraction of the total  vev \cite{EG,S}. See also \cite{DEQ} where the usual particle Higgs mixes with a SM neutral unparticle continuum.}   
This is the unparticle behavior \cite{G}, hence the name {\em unparticle Higgs}, or {\em Unhiggs} in short.
Much as the SM Higgs, the Unhiggs can develop a vev that breaks electroweak symmetry.   

The Unhiggs physics is determined by several continuous parameters.   
The exponent $d$ is called the conformal dimension, and varies between $d = 1$ (the ordinary particle limit) and $d = 2$ (the upper limit in a ghost-free theory).     
The mass gap $\mu$ defines the momentum scale where the tree-level propagator develops an imaginary part,  
which means that a continuum of new degrees of freedom opens up at the scale $\mu$. 
These new degrees of freedom are thought of as excitations of an approximately conformal hidden sector charged under the electroweak group.\footnote{See \cite{GK} for microscopic realizations of unparticles.
The microscopic model for the Unhiggs (where the unparticle sector takes part in electroweak symmetry breaking) has not been specified to date.}    
Finally, the Unhiggs is also characterized by the mass parameter $m_{uh}$ that, for momenta below the mass gap, plays a similar role as the SM Higgs mass.   

The concept of the Unhiggs implies profound modifications of the SM interactions, in particular, vertices with an arbitrary number of gauge bosons appear at the tree level. 
Nevertheless, it was demonstrated \cite{ST} that the longitudinal gauge boson scattering remains unitary, even though the analytical structure of the amplitude is completely different than in the SM.
Thus, the Unhiggs can take over one important task that in the SM is fulfilled by the Higgs boson.   
In this  paper we show that it can fulfill another task: 
it contributes to electroweak precision observables in a way consistent with the electroweak precision tests.
In fact, we will show that the Unhiggs closely mimics the SM Higgs in a large portion of its parameter space.    
While that statement is hardly surprising when the Unhiggs conformal dimension is close to $d=1$, or if the mass gap is much larger than the weak scale, it remains true also for the dimension close to $d=2$ and the mass gap smaller than 100 GeV.     

In order to compute the electroweak precision observables, we will use the holographic formulation of the Unhiggs developed in ref.\ \cite{FV_hu}.  
A non-local action can be represented as a boundary effective action of a local 5D gauge theory in a warped background.
The continuum  of excitations can be achieved in the soft-wall set-up \cite{KKSS} where the IR brane is sent away to infinity (in conformal coordinates) and the warp factor decays exponentially in the IR.  
The underlying local formulation greatly simplifies the computation of amplitudes involving the Unhiggs. 
In particular, gauge invariance is trivially realized, whereas it is a nightmare in the original non-local formulation \cite{gaugeinv}.   
Furthermore, the holographic formulation helps addressing some theoretical problems and consistency issues, e.g. fine-tuning, or the necessity for a cut-off scale.    

In the holographic approach, the Unhiggs is represented by a 5D bulk scalar field $H(x,z)$. 
The fifth dimension is warped with a warp factor that, in conformal coordinates, decays exponentially in the IR: $a(z) \sim e^{-2 \mu z/3}$ for $\mu z \gg 1$  \cite{CMT}. 
At small $z$, the 5th dimension is truncated by the  UV brane located at $z = R$ (we assume $\mu R \ll 1$), and the scale $\Lambda = 1/R$ sets  the UV cut-off scale for the Unhiggs scenario.  
There is no IR brane, and the fifth coordinate extends all the way to $z = \infty$. 
This set-up has a continuum of bulk Higgs KK excitations, but there is a mass gap $\mu$ thanks to the exponential decay of the warp factor in the IR. 

The Higgs field has a local potential in the bulk and/or on the UV brane, which triggers its vev, 
$\la H \ra = \bvec 0 \\ \hat v(z) \evec /\sqrt{2 R}$. 
The vev breaks electroweak symmetry giving masses to the SM W and Z bosons, where the electroweak scale is given by the integral of the bulk Higgs vev: $v^2 = R^{-1} \int a^3 \hat v^2$.  
We define the 5D scalar field $h$ describing the excitations in the direction of the bulk vev: 
$\hat v(z) \to \hat v(z) + h(x,z)$.
This 5D Higgs boson interacts with the SM gauge bosons and fermions, and all their KK modes. 
The gauge interactions take place in the bulk,  
since at least the $SU(2)_L \times U(1)_Y$ gauge bosons must propagate in the bulk.\footnote{If we want to preserve the custodial symmetry we need to extend $U(1)_Y$ to $SU(2)_R \times U(1)_X$ \cite{ADMS}, but this is not necessary for the sake of this paper. Furthermore, if the SM quarks propagate in the bulk then also $SU(3)$ color should  live in the bulk.}
Once the Higgs field gets a vev, the 5D Higgs boson has a single vertex with two 5D W bosons, and the coupling strength varies along the 5th dimension as $g_{L*}^2 a^3 (z)\hat v(z)/4$.  
There is also a similar coupling to two neutral fields (bulk Z bosons) with the coupling strength $(g_{L*}^2 + g_{Y*}^2)a^3 (z)\hat v(z)/4$.   
In this paper we assume that all SM fermions live on the UV boundary at $z = R$ so that their Yukawa interactions with the bulk Higgs are confined to the UV boundary: 
$\delta(z- R) R^{1/2} Y_{f*} H \bar f f$. 

Practical computations in 5D are most conveniently done using propagators defined in the mixed 4D momentum space/5thD position space.
These propagators describe the amplitude to propagate the field carrying 4D momentum $p$ from point $z$ to $z' \geq z$. 
For the Higgs boson field $h$, the mixed momentum/position propagator can be succinctly written as 
\beq
\label{e.hp}
P (p^2,z,z') = 
 {K(z,p^2) K(z',p^2) \over \Pi(p^2) }  -   R \, S(z,p^2)  K(z',p^2). 
\eeq  
Here we have introduced $K(z,p^2)$ as the solution to the bulk equation of motion that is regular in the IR (that is to say, for Euclidean $p^2 = - p_E^2 < 0$ it is damped as $e^{- p_E z}$ for  $z \to \infty$) and normalized to $K(R) = 1$.  
$S(z,p^2)$ is another independent solution that satisfies $S(R)= 0$, $S'(R) = 1$.  
$\Pi(p^2)$ is the kinetic function defined by 
\beq
\label{e.kf} 
\Pi(p^2) = R^{-1} K'(R,p^2)  + \Pi_{UV}(p^2),   
\eeq 
where $\Pi_{UV}(p^2)$ is an arbitrary local polynomial determined by the Higgs UV boundary terms. 

In a similar fashion one could write down the propagators for the gauge fields. 
In this paper we investigate the set-up where all KK excitations of the gauge fields have a very large mass gap, in 
which case all vector propagators collapse to zero-mode propagators, e.g. $P_{WW}(p^2,z,z') \to L^{-1}/(p^2 - m_W^2)$, $L= \int_R^\infty a(z)$. 
One should stress that this is {\em not} a generic situation.    
Generically, the vector mass gap would be of the same order of magnitude as that of the Unhiggs (in fact, three times lower in the simplest setting \cite{FV_hu}). 
However, the parameter space of the 5D model has corners where the Unhiggs and the vector mass gaps are separated.
This can be achieved either by means of fine-tuning or by allowing the Higgs and gauge kinetic terms to propagate in a different effective 5D metric.
In those cases, the Unhiggs mass gap can be tuned to be small, of the order of the weak scale, while the vector resonances can be much heavier and decouple from the low energy physics.   
Similar comments apply to the charged and neutral physical scalars hosted by the bulk Higgs field: naturally they would appear at the same scale as the Unhiggs, but they can be parametrically decoupled. 
In this paper we restrict to that (somewhat unnatural) simplified set-up where the only new degrees of freedom at the weak scale are those belonging to the Unhiggs continuum. 
This will allow us to zoom in on the peculiar properties of the Unhiggs physics.
We postpone the study of the more general case to a subsequent publication.  

 
The connection to the 4D non-local formulation of the Unhiggs proceeds via the boundary effective action \cite{BPR}.  
Once the SM fermions are localized on the UV brane, it is convenient to identify the SM gauge fields with the boundary value of the 5D gauge fields. 
Thus, the $SU(2)_L \times U(1)_Y$ gauge bosons are defined as $\bar L_\mu^a = L^{-1/2} L_\mu(x,R)$, $\bar B_\mu^a = L^{-1/2} B_\mu(x,R)$.
This definition ensures that the gauge fields couple to the SM fermion in a canonical way (that is, there are no vertex corrections at tree level).  
The SM gauge couplings are identified as $g_L = g_{L*} (R/L)^{1/2}$, $g_Y = g_{Y*} (R/L)^{1/2}$.    
Similarly, the Unhiggs field is the boundary value of the 5D Higgs boson field: $\bar h =  h(x,R)$. 
The quadratic part of the effective action for the Unhiggs field is the inverse of the UV-boundary-to-boundary propagator.
From \eref{hp} it follows that
\beq
\label{e.seff}
S_{\rm eff} =   \int {d^4 p \over (2 \pi)^4}  \frac{1}{2} \bar h(-p)  \Pi(p^2)  \bar h(p)   + \dots \ . 
\eeq    
The kinetic function $\Pi(s)$ in 5D warped models can have a non-trivial analytic structure in the complex $s$ plane.  
In the soft-wall scenario with the warp factor exponentially decaying in the IR, $\Pi(s)$ has a branch cut for $s > \mu^2$.
The corresponding spectral function $\rho(s) = - \pi^{-1} {\rm Im}[1/\Pi(s + i \eps)]$ vanishes below $\mu^2$, up to a possible isolated pole whose presence depends on the mass parameters in the theory, and it is positive and continuous above $\mu^2$. 
Furthermore, the boundary effective action is conformal in the UV if the warp factor is approximately that of AdS space in the vicinity of the UV brane.
In that case, $\rho(s)$ displays the unparticle power-law behavior for $R^{-2} \gg s \gg \mu^2$.  
One could use the effective action \erefn{seff} to derive predictions for the Unhiggs physics, however the underlying 5D local theory is far more convenient for practical computations.

\begin{figure}[tb]
\begin{center}
\includegraphics[width=0.5\textwidth]{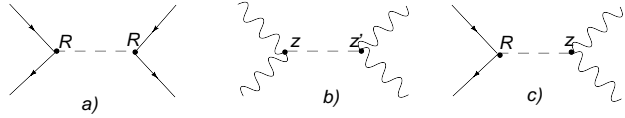}
\caption{Processes involving the Unhiggs exchange between a) two SM fermions, b) two SM gauge bosons, c) between a SM fermion and a SM gauge boson.}
\label{f.ep}
\end{center}
\end{figure}

After all these preliminary technicalities, computing Unhiggs amplitudes becomes straightforward.  
Consider first a process that involves an Unhiggs exchange between two SM fermions, fig. \ref{f.ep}a.
Since the fermions are assumed to reside on the UV boundary, both interaction vertices are on the boundary, so that the 5D Higgs boson has to propagate from the UV boundary into the bulk and back.   
The relevant propagator is thus the UV-boundary-to-boundary propagator, which is the inverse of the kinetic function $\Pi(p^2)$.
Note also that the Higgs-fermion vertex is proportional to the 5D Yukawa coupling which is related to the fermion masses by 
$Y_{f*} = \sqrt{2} m_f/\hat v(R)$. Since $\hat v(R) \neq v$ in general, the coupling strength is different than the SM one $m_f/v$. 
It is convenient to include that coupling strength rescaling in the definition of an effective propagator: 
\beq
P_{\rm eff}^{[ff]}(p^2) =   {\hat v(R)^2 \over v^2}{1 \over \Pi(p^2)}.      
\eeq 
For example, the s-wave fermion scattering amplitude reads ${m_{f}^2 \over v^2} P_{\rm eff}^{[ff]}(s)$, and differs from the corresponding SM expression by replacing the Higgs propagator $1/(s - m_h^2)$ with $P_{\rm eff}^{[ff]}(s)$.    

Next, consider a less trivial case where the Unhiggs propagates between two gauge boson vertices, fig \ref{f.ep}b.
Now both vertices can be anywhere in the bulk, so that the amplitude involves a double integral over the 5th dimension. 
Moreover, the coupling strength depends on the position and varies as $\sim a^3 (z)\hat v(z)$. 
Putting this all together, we define the effective propagator relevant for this class of processes     
\beq
P_{\rm eff}^{[gg]}(p^2) = {2 \int_R^\infty dz' \int_R^{z'} dz a^3(z) \hat v(z) a^3(z') \hat v(z')  P(p^2,z,z') \over R v^2}
\eeq	
As in the previous case, the Unhiggs amplitudes are obtained by replacing with $P_{\rm eff}^{[gg]}$ the Higgs propagator in the corresponding SM amplitude. 
For example, the Unhiggs contribution to the scattering amplitude of isospin gauge boson states $ab \to cd$ (in the limit $g_Y\to 0$ for simplicity) is given by $\delta_{ab}\delta_{cd} M(s) + \delta_{ac}\delta_{bd} M(t) + \delta_{ad}\delta_{bc} M(u)$ where    
$M(s) = - {g_L^2 \over m_W^2} [\eps(k_1) \cdot \eps(k_2)][\eps(k_3) \cdot \eps(k_4)]  P_{\rm eff}^{[gg]}(s)$ and $\eps(k)$ are the polarization vectors of the gauge bosons.  
Furthermore, $P_{\rm eff}^{[gg]}$ is the relevant propagator to compute the Unhiggs production via Higgstrahlung with a subsequent decay into WW, ZZ or $\gamma\gamma$.  

Finally, by the same logic one can define the effective propagator relevant for processes with the Unhiggs propagating between SM  fermions and gauge bosons:   
\beq 
P_{\rm eff}^{[gf]}(p^2) =  {\int_R^{\infty} a^3(z) \hat v(z)  K(z,p^2) \over R \hat v(R) \Pi(p^2)}.
\eeq  
This propagator would be relevant, for example, for Unhiggs production via Higgstrahlung with a subsequent decay into a pair of SM fermions.
 
So far our discussion has been completely general, provided the vector resonances can be decoupled. 
In particular, it is valid for arbitrary 5D backgrounds that may or may not describe approximately conformal unparticles.  
In the remainder of the paper, we will concentrate on the specific background introduced in \cite{FV_hu}, which describes the Unhiggs studied  by Stancato and Terning \cite{ST}.
The corresponding 5D background has the warp factor  $a(z) = {R \over z} e^{- 2\mu(z-R)/3}$, and the position dependent bulk Higgs mass term  $\hat M^2 (z) = {\nu^2 - 4 \over z^2} - {3 \mu \over z}$. 
In this background, the bulk Higgs vev is given by \cite{FV_hu} 
\beq
\label{e.stvs} 
\hat v(z) = v_0  a^{-3/2}(z) {z^{1/2} K_\nu(\mu z) \over R^{1/2} K_\nu(\mu R)}. 
\eeq  
This implies that the electroweak scale can be expressed as  
\beq 
v^2 =   Z_{\rm uh} v_0^2 , 
\qquad \qquad  Z_{\rm uh}= {1\over 2} \left [ {K_{1 + \nu}(\mu R) K_{1 - \nu}(\mu R) \over K_{\nu}(\mu R)^2} - 1 \right ].   
\eeq 
The Higgs equations of motion have the following two independent solutions \cite{FV_hu}:  
\bea
\label{e.ste} 
S (z,p^2) &=&  R a^{-3/2}(z) (z/R)^{1/2}
\nn && 
\cdot \left [K_\nu(\sqrt{\mu^2 - p^2} R) I_\nu(\sqrt{\mu^2 - p^2}z) - I_\nu(\sqrt{\mu^2 - p^2} R) K_\nu(\sqrt{\mu^2 - p^2}z)
\right ],
\nn
K (z,p^2) &=& a^{-3/2}(z) (z/R)^{1/2}{K_\nu(\sqrt{\mu^2 - p^2}z) \over K_\nu(\sqrt{\mu^2 - p^2} R) } .   
\eea 
Plugging the last expression into the definition \erefn{kf} of the kinetic function one obtains 
$\Pi(p^2) = \Pi_0(p^2) - Z_{\rm uh} m_{\rm uh}^2$ where  
\beq 
\label{e.stk0} 
\Pi_0(p^2) = 
 {\mu  K_{1 - \nu}(\mu R) \over R K_{\nu}(\mu R)} 
-  {\sqrt{\mu^2 - p^2} K_{1 - \nu}(\sqrt{\mu^2 - p^2} R) \over R K_{\nu}(\sqrt{\mu^2 - p^2} R) },  
\eeq 
and the Unhiggs mass parameter $m_{\rm uh}$ is a combination of the UV boundary mass and the bulk parameters $\mu,\nu, R$ \cite{FV_hu}. 
Fine-tuning various contributions we can arrive at $m_{\rm uh}$ of the order of the weak scale.  
The hierarchy problem is not addressed in this set-up.  

\begin{figure}[tb]
\begin{center}
\includegraphics[width=0.3\textwidth]{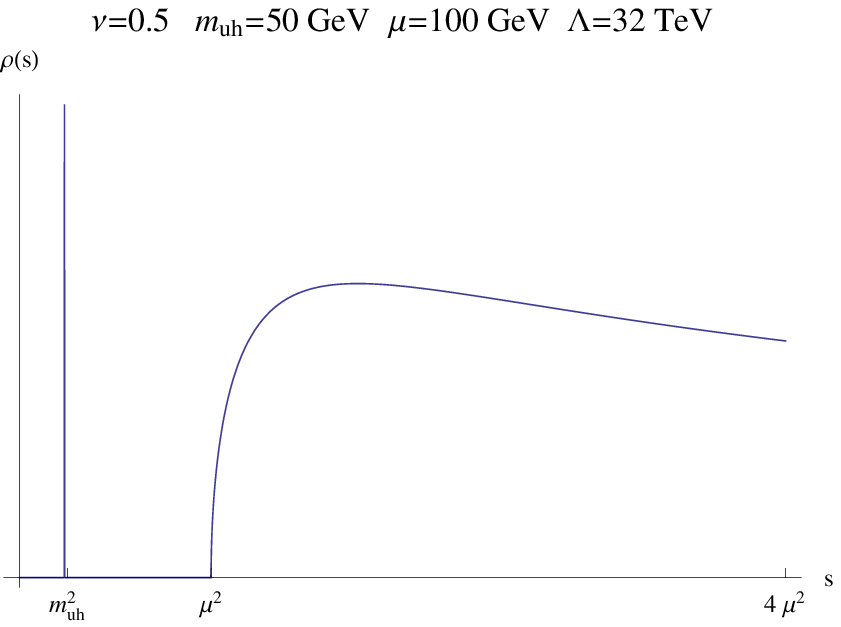}
\includegraphics[width=0.3\textwidth]{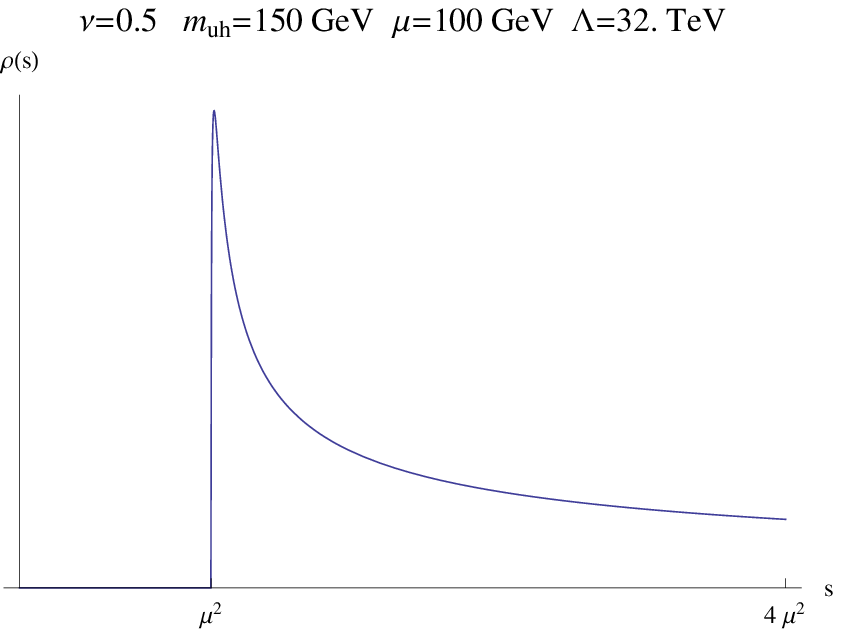}
\includegraphics[width=0.3\textwidth]{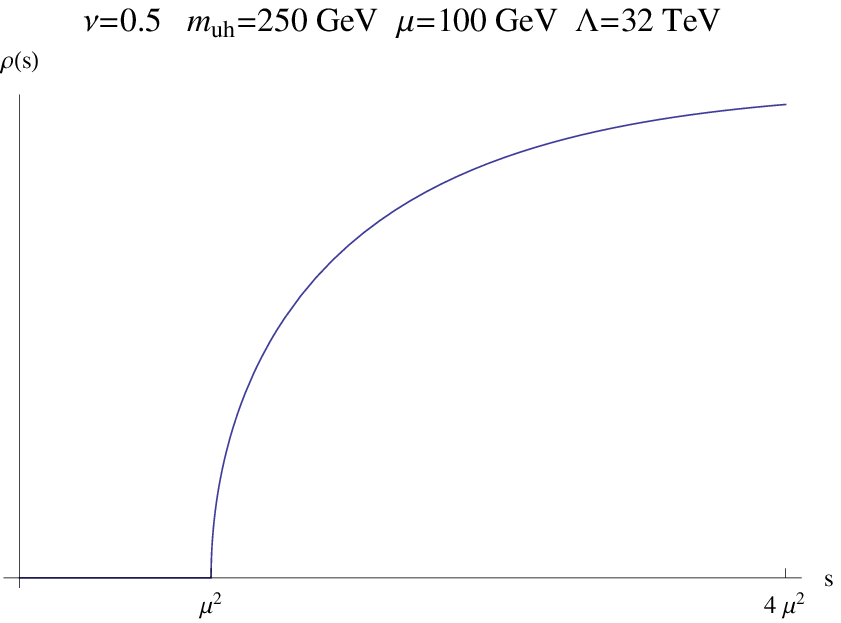}
\end{center}
\caption{Examples of the spectral function illustrating three basic types of behavior. 
In the left panel, the Unhiggs mass parameter $m_{\rm uh}$ is much smaller than the mass gap $\mu$, which results in the continuum above $s = \mu^2$ being accompanied by an isolated delta function corresponding to a particle with mass $\sim m_{\rm uh}$.
As the ratio  $m_{\rm uh}/\mu$ is increased, the isolated pole approaches the continuum and eventually merges into it (for $\nu > 0$ the merging occurs for $m_{\rm uh} \sim \nu^{1/2} \mu$).  
In the middle panel, where $m_{\rm uh} \simgt \mu$, the pole has merged with the continuum but the spectral function is peaked near $s = \mu^2$.  
In the right panel, where $m_{\rm uh} \gg \mu$, there is no sharp features.  
}
\label{f.rho}
\end{figure}

For momenta well below the mass gap, $|p^2| \ll \mu^2$, the kinetic function reduces to the normal particle kinetic function 
$\Pi(p^2) \approx Z_{\rm uh} (p^2 - m_{\rm uh}^2)$. 
Quite intuitively, the non-local character of the theory becomes manifest when $p^2 \sim \mu$, that is when the bulk degrees of freedom become kinematically available. 
For $0 < \nu < 1$ and $|p^2 R^2| \ll 1$ the kinetic function is approximated by 
$\Pi_0(p^2) \sim \mu^{2\nu} -(\mu^2 - p^2)^\nu$, which is the kinetic function assumed by Stancato and Terning after identifying $d = 2 - \nu$.
Note that the 5D model makes perfect sense also for positive $\nu$ outside the open interval $0 < \nu < 1$. 
The case $\nu > 1$ is in practice indistinguishable from the particle limit $d = 1$ and is of little interest. 
The limit $\nu = 0$ is much more interesting since it defines a smooth continuation of the unparticle dimension to $d = 2$.
 
The inverse of the UV brane position, $\Lambda = 1/R$, is the cut-off scale for the Unhiggs scenario.  
Formally, the Unhiggs of ref.\ \cite{ST} is recovered in the limit $\Lambda \to \infty$. 
Note, however, that from the 5D point of view we cannot send $\Lambda \to \infty$ without destroying perturbativity. 
Indeed, the 5D top Yukawa coupling is given by $Y_{t*} = \sqrt{2} m_{t}/v_0 \approx v/v_0 = Z_{\rm uh}^{1/2}$. 
For $0 < \nu < 1$ we find $Z_{\rm uh} \sim (\mu R)^{2\nu - 2}$ which yields  $Y_{t*} \sim (\mu R)^{\nu - 1}$. 
Perturbativity requires $Y_{t*} \simlt 4 \pi$ which sets the upper limit on the cut-off scale $\Lambda \sim  \mu (4\pi)^{1/(1-\nu)}$ (in practice, this relation is modified by a $\nu$-dependent numerical factor of order few). 
In particular, for $\nu$ approaching $0$ (the conformal dimension approaching $2$), the  cut-off is parametrically $\Lambda \sim 4 \pi \mu$.      

Next, plugging the solutions \erefn{ste} into our general formulas for the effective propagators we obtain  
\bea
\label{e.steup} 
P_{\rm eff}^{[ff]}(p^2) &=& {Z_{\rm uh} \over  \Pi_0(p^2) - Z_{\rm uh} m_{\rm uh}^2 + i Z_{\rm uh} m_h \Gamma_h(p^2) },
\nn
P_{\rm eff}^{[gg]}(p^2) &=& {1 \over p^2}  + {\Pi_0(p^2) \over p^4}  {m_{\rm uh}^2 - i m_h \Gamma_h(p^2) \over \Pi_0(p^2) - Z_{\rm uh} m_{\rm uh}^2 + i Z_{\rm uh} m_h \Gamma_h(p^2)}, 
\nn
P_{\rm eff}^{[gf]}(p^2) &=&  {\Pi_0(p^2) \over p^2} {1 \over  \Pi_0(p^2) - Z_{\rm uh} m_{\rm uh}^2 + i Z_{\rm uh} m_h \Gamma_h(p^2) }.
\eea
Above, we included the fermionic width which is relevant for production processes.  
The fermionic width amounts to shifting $\Pi(p^2) \to \Pi(p^2) + i Z_{uh} m_{h} \Gamma_h(p^2)$ in the denominator of the boundary part of the 5D  propagator \erefn{hp}, where $m_h \Gamma_h(p^2) \approx {3 \over 8 \pi} {m_b^2 \over v^2} p^2$ is the $H \to b \bar b$ decay width of the SM Higgs boson, and the factor $Z_{uh}$ accounts for the modified coupling of the Unhiggs to the UV boundary fermions.   
In each case, for $p^2 \ll \mu^2$ we get  $P_{\rm eff} \approx 1/(p^2 - m_{\rm uh}^2 + i m_h \Gamma_h(p^2))$:
far below the mass gap the effective propagators reduce to the normal Higgs propagators, and $m_{\rm uh}$ can be identified with the Higgs mass.   
For large $p^2$, such that $\Pi_0(p^2) > Z_{\rm uh} m_{\rm uh}^2$, the leading behavior of the gauge-gauge and the gauge-fermion propagators is  $P_{\rm eff} \approx 1/p^2$.
This means that the amplitudes involving the electroweak gauge bosons have the same UV properties as in the SM.  
This important observation ensures perturbative unitarity of the longitudinal gauge boson scattering (and also fermion-antifermion scattering into the longitudinal gauge bosons).  
The $1/p^2$ UV asymptotics will also be essential to demonstrate that the loop corrections to electroweak precision observables are finite. 
The fact that the UV asymptotics is analytic, despite the Unhiggs ``kinetic" term in the effective action being non-analytic, can be traced to the fact that the Higgs-gauge vertex is smeared in the bulk.
On the other hand, the fermion-Higgs vertex is localized on the UV brane and the effective fermion-fermion propagator displays a non-analytic UV behavior. 

Note also that, for $m_{\rm uh} \to 0$, $P_{\rm eff}^{[gg]}$ and $P_{\rm eff}^{[gf]}$ are equal to $1/p^2$ (ignoring the width), because in this limit the ``KK modes"  of the bulk Higgs are orthogonal to the zero modes of W and Z gauge bosons. In the opposite limit,  $m_{\rm uh} \to \infty$, the fermion-fermion and the gauge-fermion propagators vanish. 
That is because this limit implies taking the brane Higgs mass to infinity, which pushes the bulk Higgs away from the UV brane where the fermions live.

\begin{figure}[tb]
\begin{center}
\includegraphics[width=0.3\textwidth]{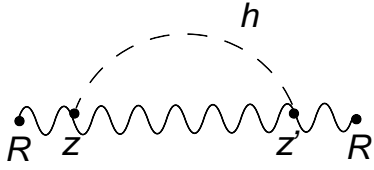}
\caption{The 5D diagram relevant for the Unhiggs contribution to S and T.}
\label{f.std}
\end{center}
\end{figure}

We are ready to address the central question of this paper: what is the impact of the Unhiggs on electroweak precision observables? 
Much as in the SM, we need to compute the vacuum polarization diagrams of the electroweak gauge bosons.
More precisely, in the 5D holographic approach we need to compute one-loop Unhiggs corrections to the UV-boundary-to-boundary propagator depicted in \fref{std}.   
As we explained, in the limit where the vector resonances are decoupled, this boils down to replacing the Higgs propagator in the corresponding SM amplitude with the relevant effective Unhiggs propagator. 
In the case at hand we need the gauge-gauge effective propagator.
This leads to the following expressions for the Unhiggs contributions to the S and T parameters: 
\bea 
\label{e.stst} 
T_{\rm uh} & = &  - {3 \over 8 \pi \cos^2 \theta_W} \int dk  {k^5 \over (k^2 + m_W^2)(k^2 + m_Z^2)}  P_{\rm eff}^{[gg]}(-k^2),
\nn 
S_{\rm uh} & = &  {1 \over 6 \pi} \int dk k^3  {k^4 +  3 k^2 m_Z^2 + 12 m_Z^4   \over (k^2 +  m_Z^2)^3} P_{\rm eff}^{[gg]}(-k^2).  
\eea  
In the SM, the Higgs contribution is logarithmically divergent, 
$T_{\mathrm{SM}}(m_h) \approx {3 \over 8 \pi \cos^2 \theta_W} \log(\Lambda/m_h)$,  
$S_{\mathrm{SM}}(m_h) \approx - {1 \over 6 \pi} \log(\Lambda/m_h)$. 
That divergence is canceled by the loops involving the electroweak gauge bosons only. 
Rather than displaying the full result, it is more convenient to define the shift of S and T with respect to some reference Higgs mass. 
Such quantities are finite in the SM: $(\Delta T)_{\mathrm{SM}}  \approx - {3 \over 8 \pi \cos^2 \theta_W} \log(m_h/m_{\mathrm{ref}})$,  
$(\Delta S)_{\mathrm{SM}} \approx {1 \over 6 \pi} \log(m_h/m_{\mathrm{ref}})$.  
For the Unhiggs, the fact that the UV behavior of the effective propagator is the same as in the SM implies that the cancellation of divergences works the same way as in the SM.
Thus, we can define the shift of S and T as 
\beq
\Delta T = T_{\rm uh} - T_{\mathrm{SM}}(m_{\mathrm{ref}}), 
\qquad 
\Delta S = S_{\rm uh} - S_{\mathrm{SM}}(m_{\mathrm{ref}}),  
\eeq
and we are guaranteed that these shifts are finite in the entire parameter space of the Unhiggs. 
In the following we fix $m_{\mathrm{ref}} = 100 \gev$. 

\begin{figure}[tb]
\begin{center}
\includegraphics[width=0.3\textwidth]{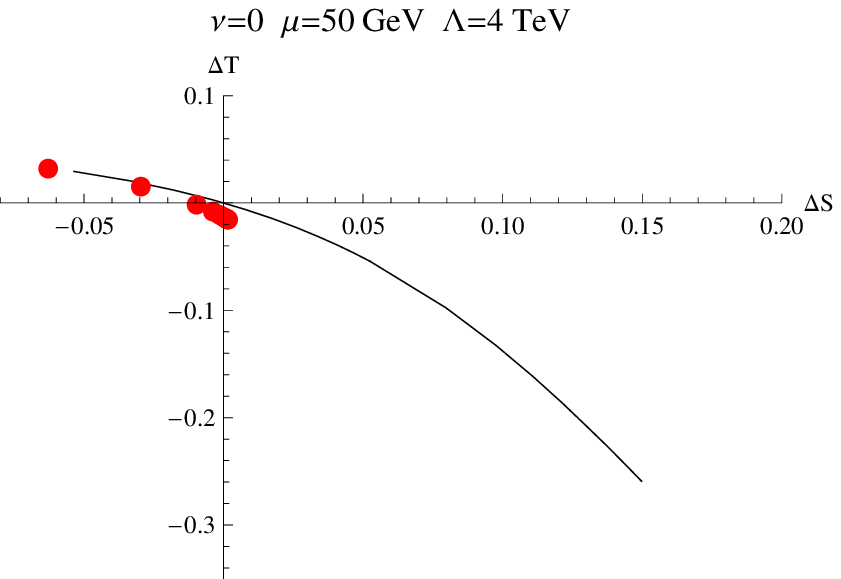}
\includegraphics[width=0.3\textwidth]{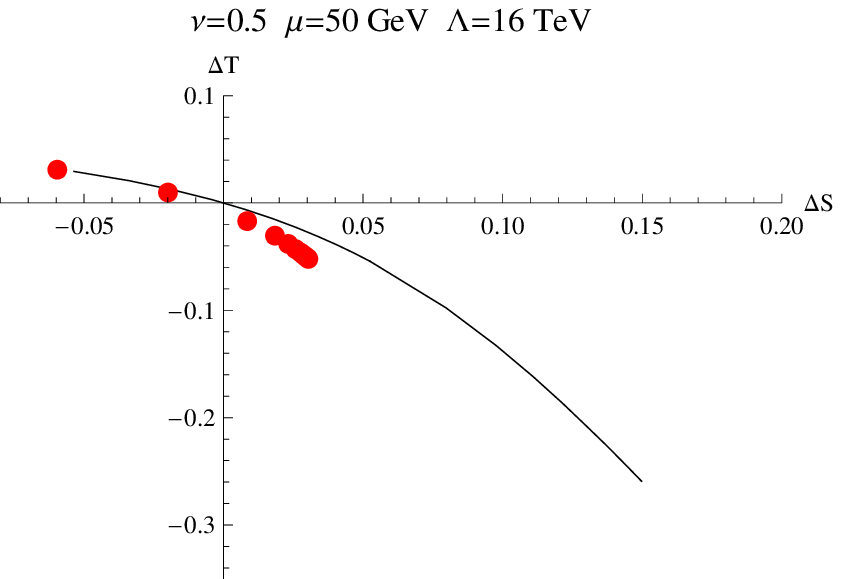}
\includegraphics[width=0.3\textwidth]{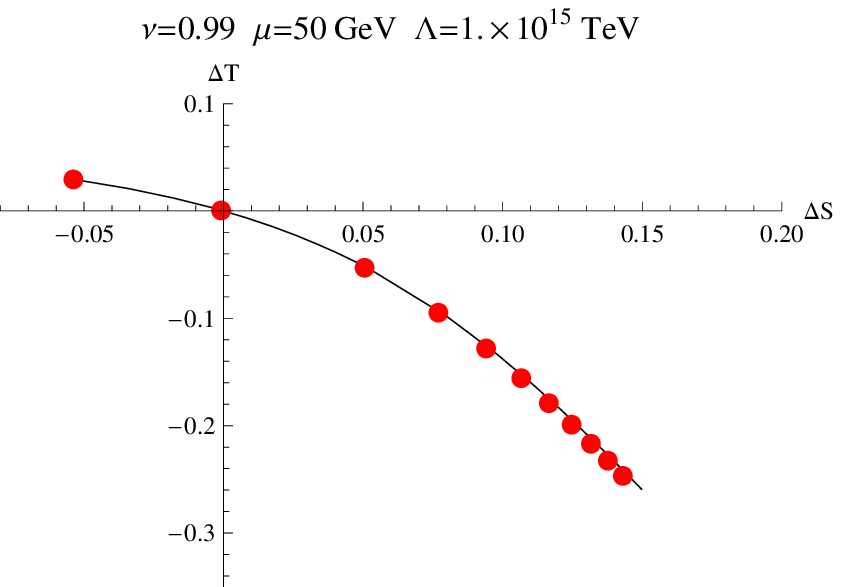}
\end{center}
\begin{center}
\includegraphics[width=0.3\textwidth]{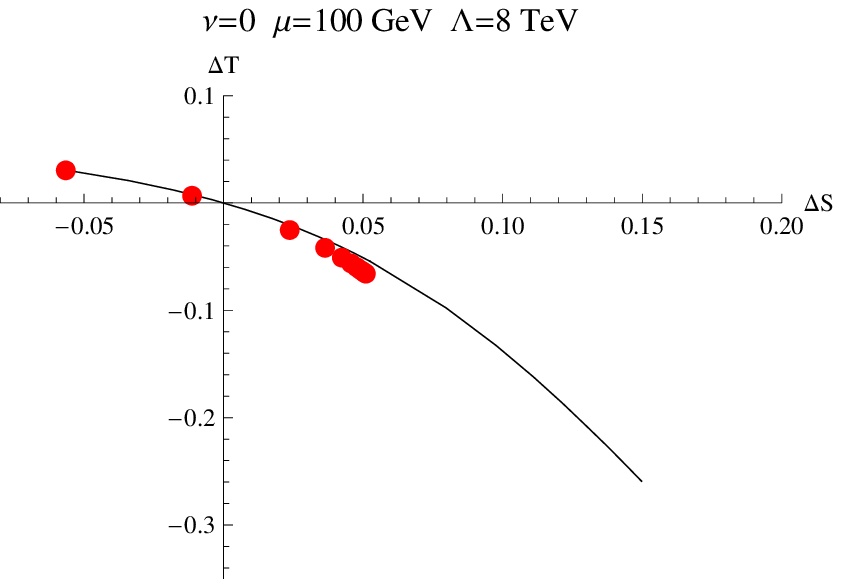}
\includegraphics[width=0.3\textwidth]{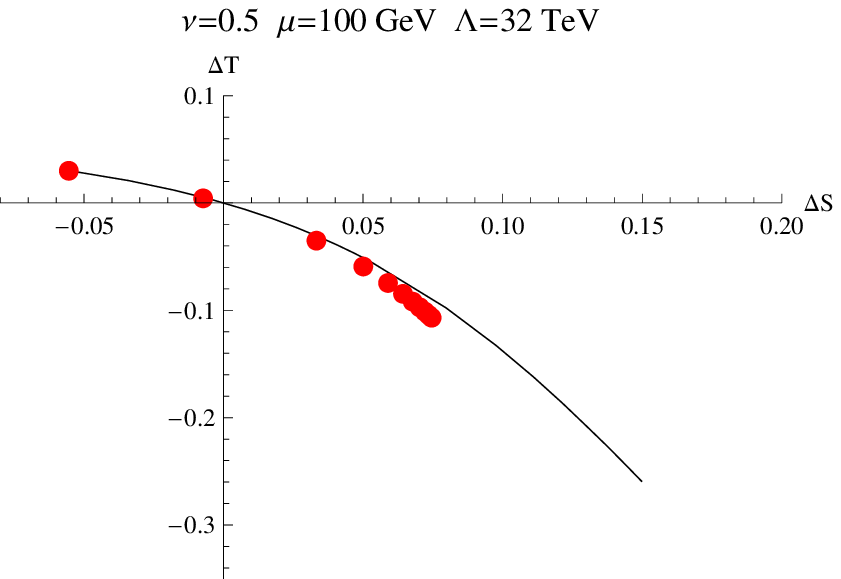}
\includegraphics[width=0.3\textwidth]{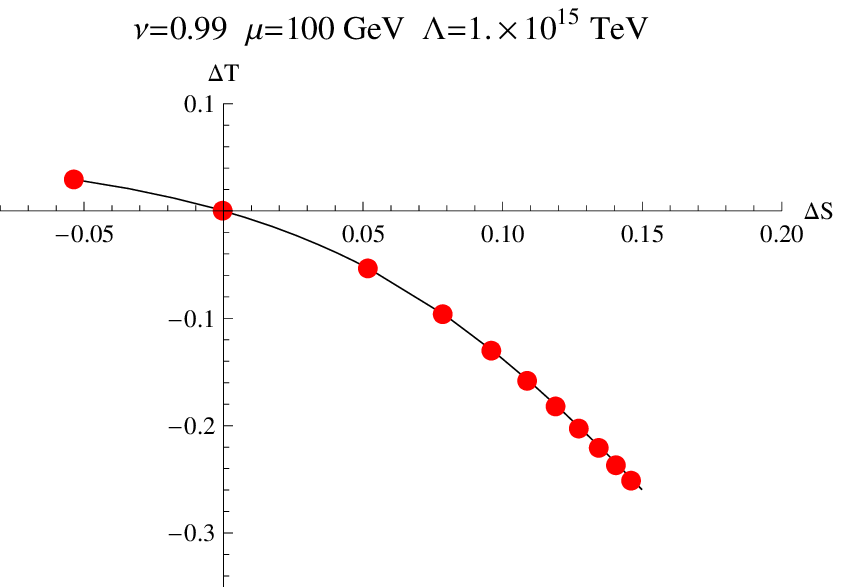}
\end{center}
\caption{
The trajectories in the $\Delta S - \Delta T$ plane for constant Unhiggs parameters $\nu$, $\mu$, $R$ and varying $m_{\rm uh}$.
For a given $\mu$ and $\nu$, the UV scale $\Lambda = 1/R$ is set to the maximum value allowed by perturbativity (or to the Planck scale, if the former is larger).  
Different points correspond to varying the Unhiggs mass $m_{\rm uh}$ in the range  $[50,1000] \gev$ (red circles, from left to right, in steps of hundred except for the first step).  
For reference, in each case we plotted the trajectory in the SM when $m_h$ is varied in the range $[50,1000] \gev$ (solid black). 
 }
\label{f.stt1}
\end{figure}
\begin{figure}[tb]
\begin{center}
\includegraphics[width=0.3\textwidth]{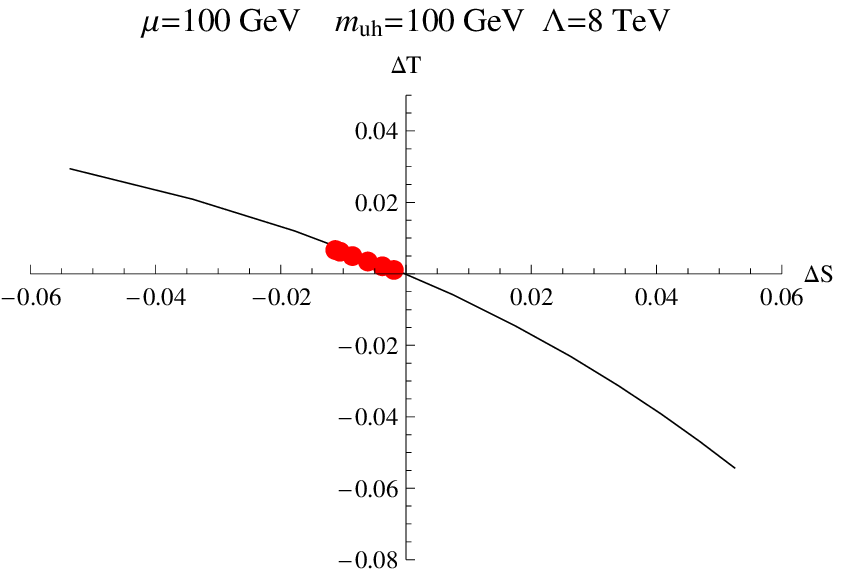}
\includegraphics[width=0.3\textwidth]{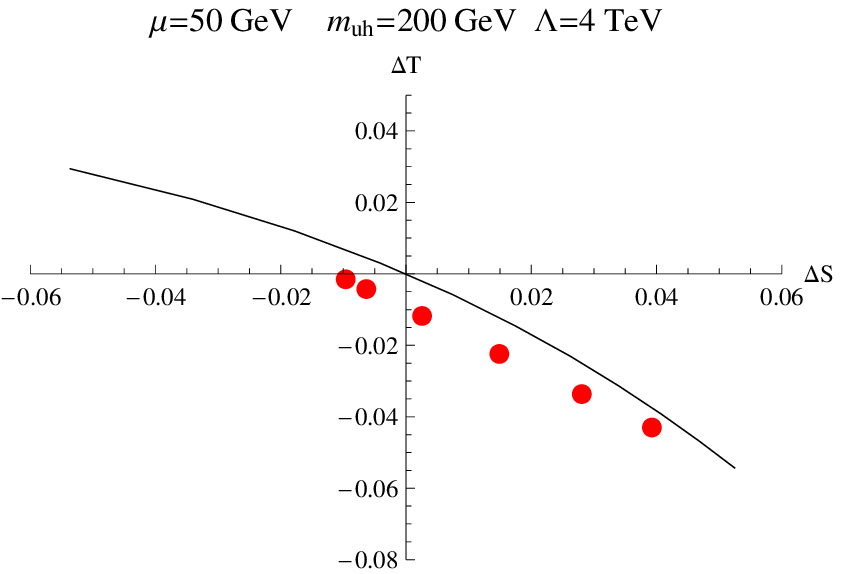}
\includegraphics[width=0.3\textwidth]{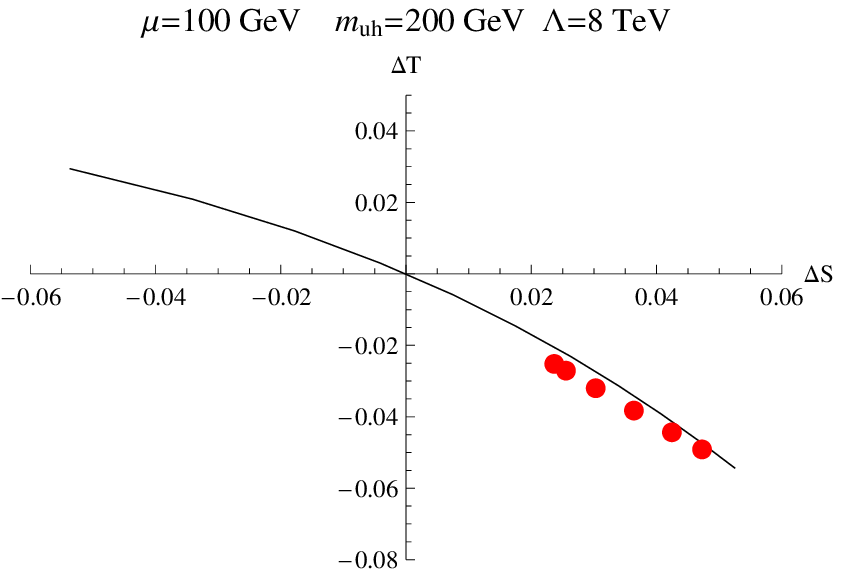}
\end{center}
\caption{
The trajectories in the $\Delta S - \Delta T$ plane for constant Unhiggs parameters $m_{\rm uh}$, $\mu$, $R$ and varying $\nu$ in the range $[0,1]$ (red circles, from left to right, in step of $0.2$).  
For reference, we plotted the  trajectory in the SM when $m_h$ is varied in the range $[50,200] \gev$ (solid black). 
}
\label{f.stt2}
\end{figure}

For a given set of $m_{\rm uh},\mu,\nu,R$ we can easily evaluate $\Delta S$ and $\Delta T$ numerically.  
Our results for sample slices of the Unhiggs parameter space are presented in  figs. \ref{f.stt1}, \ref{f.stt2} 
and compared to the SM results.  
These numerical results lead us to the following conclusions:
\bi
\item The Unhiggs is consistent with the LEP constraints on S and T in a large portion of its parameter space. 
In particular, electroweak precision observables do not exclude the conformal dimension $d = 2$ ($\nu = 0$), 
nor a mass gap smaller than $100 \gev$.    
\item Typically, the Unhiggs mimics the SM Higgs, in the sense that its contributions to $\Delta S$ and $\Delta T$ are similar to the SM Higgs contribution for {\em some} Higgs mass.
\item 
If the Unhiggs mass parameter $m_{\rm uh}$ is much smaller than the mass gap $\mu$, then the Unhiggs mimics the SM Higgs with mass $m_h \sim m_{\rm uh}$.  
If, on the other hand, $m_{\rm uh} \gg \mu$ and $d$ is away from $1$ (the particle limit), the Unhiggs mimics the SM Higgs with mass $m_h \sim \mu$.   
The latter can be partly understood from the behavior of the spectral function $\rho(s)$, which has a large support near $s = \mu^2$, independently of the value of $m_{\rm uh}$.    
\ei 
We conclude that the electroweak precision observables are consistent with the Unhiggs with a mass gap of order 100 GeV, irrespectively of whether there is an isolated pole below the continuum or not. 

What about the direct searches at LEP?   
When the Unhiggs spectral function has a pole well below the continuum (as is the case when $m_{\rm uh} \simlt \mu$), that pole behaves much like the SM Higgs and the $115$ GeV lower limit from LEP does apply.
That is because in that case the effective Unhiggs propagators reduce to the SM Higgs propagator for $p^2 \ll \mu^2$ (including $p^2 \sim m_{\rm uh}^2$, where the resonance is located).
If, on the other hand, there is no isolated pole, then the physical properties of the Unhiggs are vastly different and the LEP limits have to be reconsidered.  

\begin{figure}[tb]
\begin{center}
\includegraphics[width=0.3\textwidth]{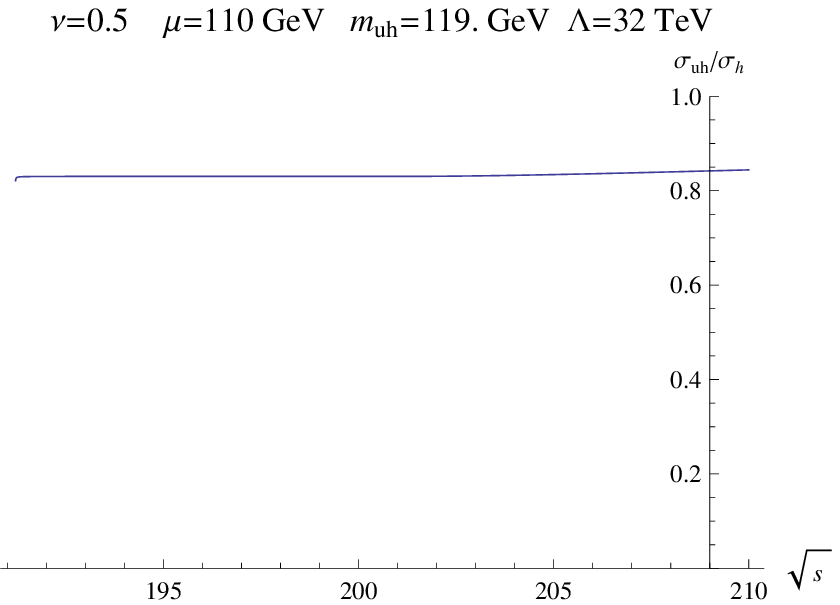}
\includegraphics[width=0.3\textwidth]{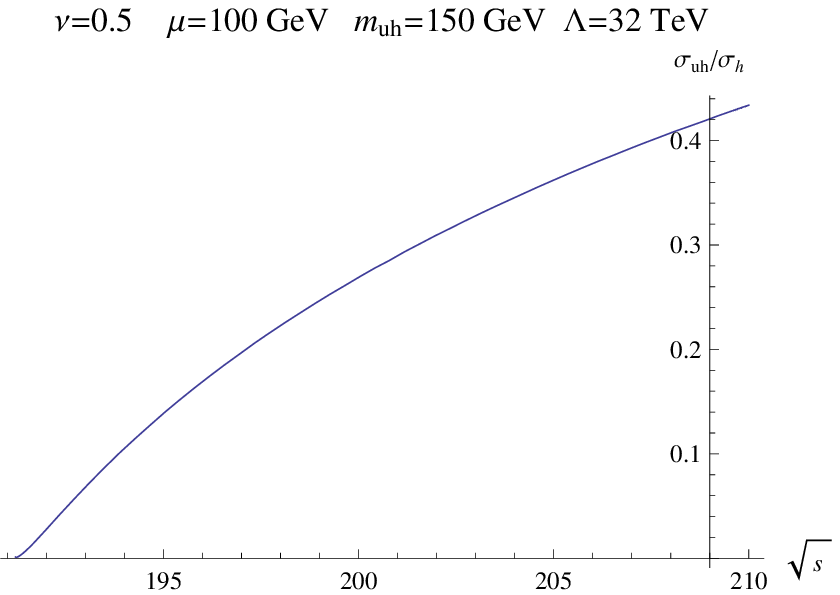}
\includegraphics[width=0.3\textwidth]{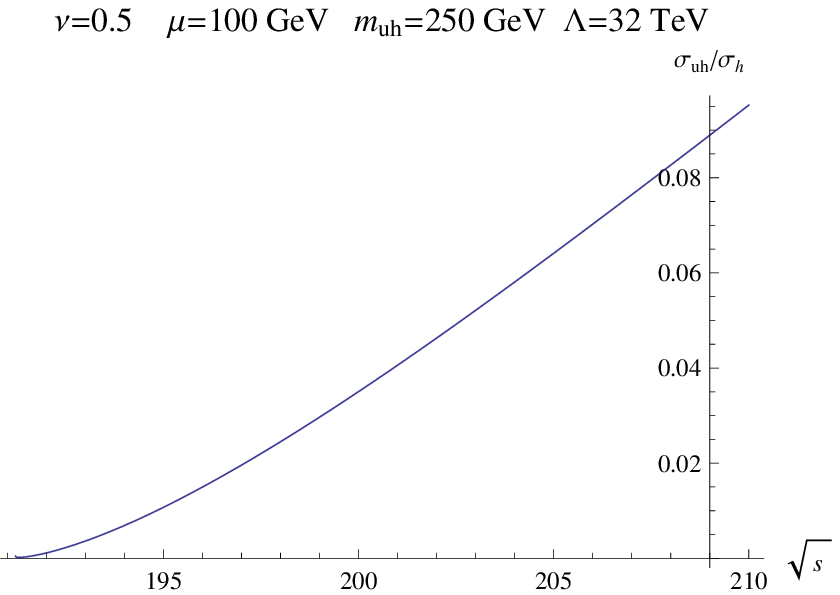}
\end{center}
\caption{
The ratio of the  total Unhiggs production cross section to the SM Higgs cross section for $m_h= 100$ GeV as a function of LEP center-of-mass energy.
In the left panel the mass parameters are chosen such that the Unhiggs propagator has an isolated pole at 100 GeV, while the continuum starts at 110 GeV. 
In the middle and right panels there is no pole below the continuum. For the choice of parameters in the middle panel the spectral function is sharply peaked near $s = \mu^2$, whereas in the right panel the spectral function is smeared out and has no peak.
}
\label{f.sigmatot}
\end{figure}

\begin{figure}[tb]
\begin{center}
\includegraphics[width=0.3\textwidth]{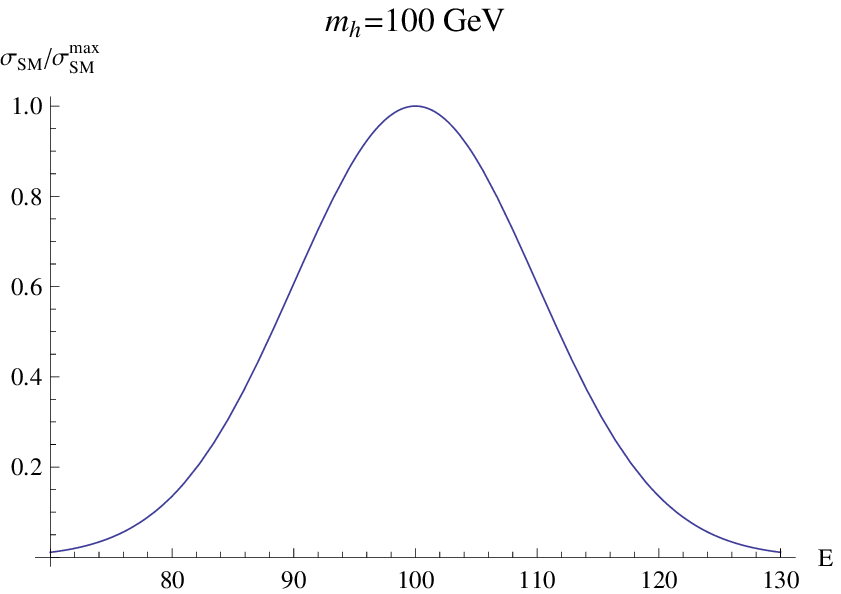}
\includegraphics[width=0.3\textwidth]{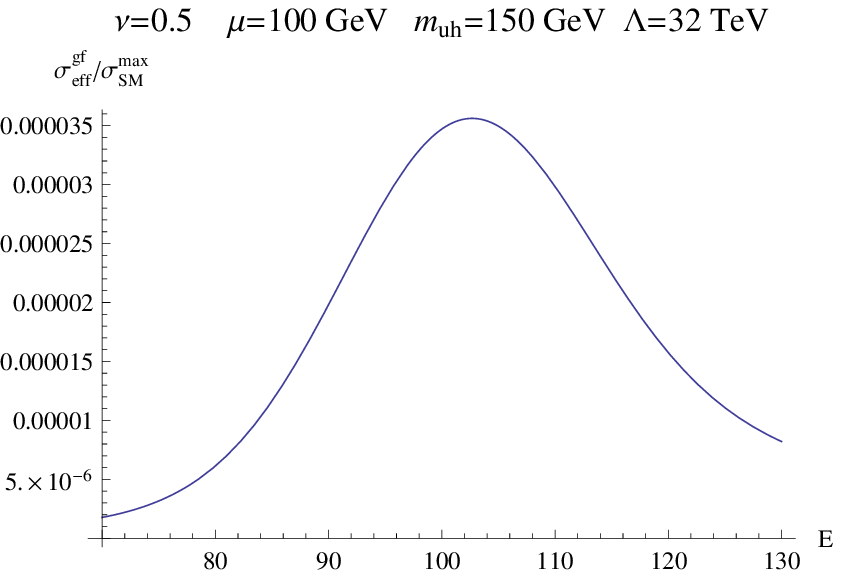}
\includegraphics[width=0.3\textwidth]{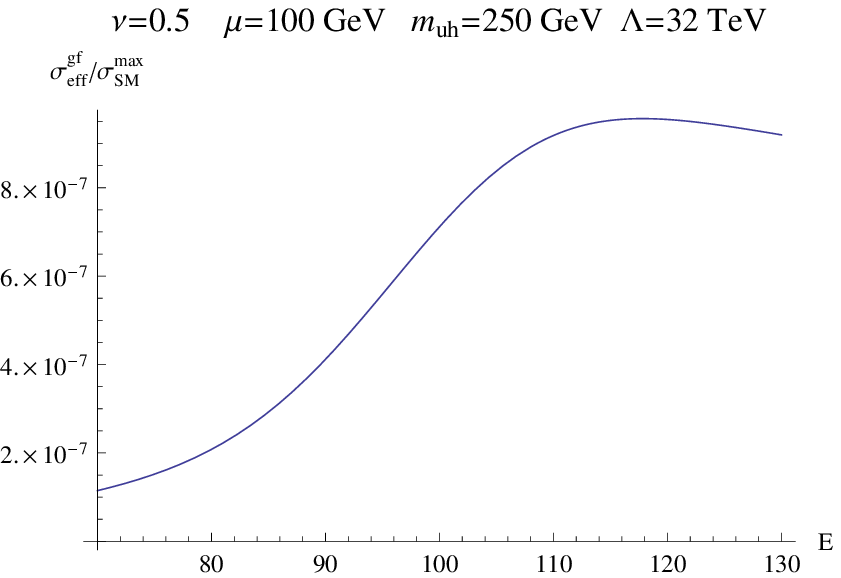}
\end{center}
\caption{
The normalized cross section for Higgs production via Higgstralung with subsequent decay into $b \bar{b}$ in the SM (left) as a function of the Higgs center-of-mass energy $E$, compared with two examples of the corresponding Unhiggs production cross-section.
In the middle panel the parameters are such that the spectral density is peaked near $s = \mu^2$, which results in the cross-section being peaked  near $E = \mu$.  
In the right panel, the spectral density is monotonically growing and the cross section is more smeared out. 
In both cases, the peak production cross section is dramatically suppressed with respect to the SM one.}
\label{f.sigma}
\end{figure}

In the SM, the cross section for the Higgs production in the Higgsstrahlung process is proportional to 
\beq
\label{e.hcs}
\sigma_{\mathrm{SM}}(E) \sim \int d \bar E f_{\sigma}(E- \bar E){m_h  \Gamma_h(\bar E{}^2) \over  (\bar E{}^2 - m_h^2)^2 + m_h^2 \Gamma_h(\bar E{}^2)^2}.
\eeq 
Here, $f_{\sigma}$ is a Gaussian distribution of width $\sigma$, which naively accounts for experimental uncertainties (we take $\sigma = 10 $ GeV). Next, $E$ is the center-of-mass energy of the emitted Higgs boson, and $\Gamma_h$ is the Higgs width. 
We focus here on the energies accesible at LEP, $E \sim 100 \gev$, in which case the latter is in practice the width of the $H \to b \bar b$ decay.
For the sake of reference, in \fref{sigma} we plotted the distribution \erefn{hcs} normalized by $\sigma_{\mathrm{SM}}^{max} = \sigma_{\mathrm{SM}}(m_h^2)$ for $m_h = 100 \gev$. The width of this distribution is set by the experimental uncertainties that we parametrize by $\sigma$. 

In the Unhiggs scenario, all the elements that enter the SM cross section are affected: the propagator, the width, the vertex with the Z boson, the vertex with the final states. 
However, the modification of the vertices is already taken into account in our definition of the effective Unhiggs propagators. 
Thus, at the end of the day, switching from the SM Higgs to the Unhiggs  boils down to replacing the Higgs propagator with the effective Unhiggs propagator. 
The {\em total} Unhiggs production cross section can be calculated by replacing the phase space of one Higgs particle $\delta(p^2- m_h^2)$ with the unparticle phase space element \cite{G,ST}, which in our language is given by $-\pi^{-1} \mathrm{Im} P_{\mathrm{eff}}^{[gg]}$.
The sample results are plotted in \fref{sigmatot}.
Suppression of the production of the Unhiggs continuum depends on the parameter space. 
When the spectral function has a pole below the continuum at $s \sim m_0^2$ then the presence of the continuum has little impact, and the production cross section ends up being similar to the SM Higgs production cross section for $m_h \sim m_0$. 
When there is no pole but the spectral function is peaked near the mass gap at $s = \mu^2$ then the production of the continuum is suppressed by a factor of $2 - 3$ compared to the SM Higgs with $m_h \sim \mu$. 
Smearing out the spectral function leads to further suppression. 
One difference with respect to the results of \cite{ST} is that we cannot switch off the production cross section by going to the limit $d \to 2$. 
In our case, the corresponding limit $\nu \to 0$ is perfectly regular, and the total production cross section only slightly differs from the $\nu = .5$ case plotted in \fref{sigmatot}.    

One can also ask the question about the {\em visible} cross section, that is about the Unhiggs production followed by the decay into asymptotic SM states.  
For energies in the LEP range only the SM fermions are kinematically available.
Therefore the Unhiggs propagation between the Z-boson vertex and the final states is described by the gauge-fermion effective propagator.  
Taking into account the Gaussian smearing (as in the SM case, \eref{hcs}), the visible cross section is proportional to  
\beq
\sigma_{\rm eff}^{[gf]}(E) \sim \int d \bar E f_{\sigma}(E- \bar E) m_h  \Gamma_h(\bar E{}^2) |P_{\mathrm{eff}}^{[gf]}(\bar E{}^2)|^2.  
\eeq  
This distribution sets the upper limit for the Unhiggs production followed by its decay within the detector. 
In \fref{sigma}, we plot this distribution for two sample points in the Unhiggs parameter space for which there is no isolated pole below the continuum.
It is clear that the visible Unhiggs cross-section is suppressed by several orders of magnitude with respect to one for the SM Higgs  with $m_h \sim \mu$. 
This makes the Unhiggs continuum 
invisible to LEP for all practical purpose.   
The relative suppression is the result of a non-resonant behavior due to the ``classical width" -- the large imaginary part of $\Pi_0(p^2)$ for $p^2 > \mu^2$,  which completely drowns the small quantum width of order $m_h \Gamma_h$. Unlike in the case of the SM Higgs, the amplitude for the Unhiggs production-plus-decay process does not pick up a strong enhancement factor from the on-shell propagator.

In summary, we have computed corrections to the S and T parameters from the Unhiggs loops.
The conclusion is that the electroweak precision observables are consistent with the Unhiggs scenario, even when the Unhiggs conformal dimension largely deviates $d = 1$, or when the Unhiggs mass gap is of order the weak scale.    
In fact, when it comes to electroweak precision observables, the Unhiggs mimics a light SM Higgs boson in a large portion of its parameter space.
At the same time, the invisible and visible Unhiggs production cross section can be suppressed relative to the SM Higgs one. 
This, together with the absence of resonant peaks,  makes the Unhiggs continuum more elusive to collider searches, in the similar spirit as in the  uniform Higgs \cite{EG} or in the stealthy Higgs \cite{BB} models. 
A more careful analysis of the LEP constraints is necessary in order to determine the allowed parameter space of the Unhiggs scenario.

\section*{Acknowledgments}

A.F. thanks Jack Gunion, Matt Strassler and John Terning for comments and discussions. 
M.P.V. is supported in part by MEC project FPA2006-05294 and Junta de Andaluc\'{\i}a projects FQM 101, FQM 00437 and FQM 03048.


\end{document}